\documentclass[12pt]{iopart}

\usepackage{gensymb}
\usepackage{textcomp}
\usepackage{graphicx}
\usepackage{subcaption}
\usepackage{hyperref}
\usepackage{float}
\usepackage{tikz}
\usepackage{siunitx}
\usepackage{xcolor}
\usepackage{textcomp}
\usepackage{tabularx}
\usepackage{graphicx}

\usepackage[margin=1in]{geometry}

\makeatletter
\renewcommand\subsection{\@startsection{subsection}{2}{\z@}%
  {-3.25ex\@plus -1ex \@minus -.2ex}%
  {1.5ex \@plus .2ex}%
  {\normalfont\normalsize\bfseries}}
\renewcommand\subsubsection{\@startsection{subsubsection}{3}{\z@}%
  {-2.5ex\@plus -1ex \@minus -.2ex}%
  {1.0ex \@plus.2ex}%
  {\normalfont\normalsize\bfseries}}
\makeatother

\begin{document}

\title[]{Novel Direct Alpha Spectroscopy Technique for $^{225}$Ac Radiopharmaceutical Detection in Cancer Cells }

\author{Mahsa Farasat$^a$, Behrad Saeedi$^{b,f}$, Luke Wharton$^b$, Sidney Shapiro$^{a,d}$, Chris Vinnick$^c$, Madison Daignault$^c$, Meghan Kostashuk$^{g,b}$, Nicholas Pranjatno$^e$, Myla Weiman$^{a,h}$, Corina Andreoiu$^c$, Hua Yang$^{b,c,f}$, Peter Kunz$^{a,c}$}
\vspace{10pt}
\address{$^a$ Accelerator Division, TRIUMF, 4004 Wesbrook Mall, Vancouver, BC V6T 2A3, Canada}
\address{$^b$ Life Sciences Division, TRIUMF, 4004 Wesbrook Mall, Vancouver, BC V6T 2A3, Canada}
\address{$^c$ Department of Chemistry, Simon Fraser University, Burnaby, BC V5A 1S6, Canada}
\address{$^d$ Physics and Astronomy, University of British Columbia, Vancouver, BC V6T 1Z4, Canada}
\address{$^e$ Department of Physics, Simon Fraser University, Burnaby, BC V5A 1S6, Canada}
\address{$^f$ Department of Chemistry, University of British Columbia, Vancouver, BC V6T 1Z1, Canada}
\address {$^g$ McMaster School of Biomedical Engineering, McMaster University, Hamilton, Ontario L8S 4L7 , Canada}
\address {$^h$ Department of Physics and Astronomy, University of Waterloo, Waterloo, Ontario N2L 3G1 , Canada}
\ead{mfarasat@triumf.ca}
\vspace{10pt}
\begin{indented}
\item

\end{indented}

\begin{abstract}
\textbf{Objective:}  
Targeted alpha-particle therapy (TAT) is a promising approach for treating metastatic cancers, utilizing alpha-emitting radionuclides conjugated to tumor-targeting molecules. Actinium-225 ($^{225}$Ac) has emerged as a clinically relevant candidate due to its decay chain, which produces four successive alpha emissions, effectively damaging cancer cells. However, the nuclear recoil effect can lead to off-target redistribution of decay daughters, complicating dosimetry and increasing potential toxicity. This study aims to address these challenges by developing a direct alpha spectroscopy method for in vitro investigations of $^{225}$Ac radiopharmaceuticals.  \textbf{Approach:}  
We developed the Bio-Sample Alpha Detector (BAD), a silicon-based detector designed to operate under ambient conditions, enabling direct alpha spectroscopy of cell samples. AR42J rat pancreatic tumor cells, which express somatostatin receptor 2 (SSTR2), were incubated with [$^{225}$Ac]Ac-crown-TATE, [$^{225}$Ac]Ac-PSMA-617, and [$^{225}$Ac]Ac$^{3+}$. The BAD setup allowed radiolabeled cell samples to be positioned within 100 $\mu$m of the detector for alpha spectra acquisition with statistical uncertainties of less than 1\% in count rates. Geant4 Monte Carlo simulations were employed to validate the experimental results. \textbf{Main Results:}  
Distinct spectral differences between radiolabeled cells and reference samples confirmed the uptake of [$^{225}$Ac]Ac-crown-TATE by AR42J cells. Detection of $^{213}$Po, a decay daughter of $^{225}$Ac, indicated partial retention and release of decay products from cells, providing insight into intracellular retention and radionuclide redistribution. Geant4 simulations confirmed the alignment of experimental data with theoretical predictions.  \textbf{Significance:}  
This study introduces a novel method for directly measuring the behavior of $^{225}$Ac and its decay daughters in biological samples using alpha spectroscopy. The BAD setup provides a valuable tool for investigating radionuclide retention, redistribution, and microdosimetry in radiopharmaceutical research. 

\end{abstract}

%
\vspace{2pc}

\section{Introduction}

The field of radiopharmaceuticals has witnessed significant advancements in recent years, particularly in the development of alpha-emitting isotopes for Targeted Alpha Therapy (TAT). TAT leverages the high linear energy transfer (LET) and short range of alpha particles to selectively destroy cancer cells while minimizing damage to surrounding healthy tissue \cite{radchenko2021production,jang2023targeted,eychenne2021overview}.This approach shows particular promise for treating metastatic and micro-metastatic cancers, which often prove resistant to conventional therapies \cite{allen2014targeted}. Alpha particles, characterized by energies typically ranging from 3 to 10 MeV, exhibit a high LET and a short range in biological tissues ($<$100 $ \mu$m). These properties enable the delivery of highly localized and lethal radiation doses, inducing severe DNA damage in targeted cells. However, these same properties pose challenges in research, as their short range complicates direct detection, particularly in biological environments where attenuation and scattering can hinder accurate measurements \cite{sollini2020five}. One alpha-emitting isotope that has gained considerable attention for its therapeutic properties is Actinium-225 ($^{225}$Ac).  $^{225}$Ac has shown significant promise for clinical applications due to its relatively long half-life of approximately 10 days. This provides a potentially suitable balance between allowing sufficient time for targeted delivery and sustaining therapeutic activity. The decay chain of $^{225}$Ac includes the net emission of four alpha particles  with energies ranging from 5.8 MeV to 8.4 MeV, each of which is highly effective in inducing double-strand breaks in the DNA of cancer cells, ultimately leading to cell death. This makes $^{225}$Ac-labeled radiopharmaceuticals particularly effective in treating cancers such as prostate cancer and neuroendocrine tumors \cite{jang2023targeted,miederer2008realizing}. 

Despite its advantages, the application of $^{225}$Ac-based therapies faces key challenges, particularly in the detection and measurement of alpha emissions in biological systems \cite{miederer2008realizing}. One significant issue is that $^{225}$Ac-labeled complexes contain up to seven radioisotopes in their decay chain. The high recoil energy from alpha decay can disrupt chemical bonds, releasing daughter radionuclides that may redistribute in vivo. Although several in vivo studies have mapped the biodistribution of key daughter isotopes such as Francium-221 ($^{221}$Fr), Astatine-217 ($^{217}$At), and Bismuth-213 ($^{213}$Bi), these approaches do not provide direct alpha particle measurements. The redistribution patterns have been shown to vary depending on the pharmacokinetics of the targeting vector, with slower-clearing antibodies releasing daughters into circulation and faster-clearing small molecules directing them toward the liver or kidneys \cite{banerjee2021preclinical, josefsson2018pharmacokinetic}. These findings have significantly improved dosimetry models and highlighted the importance of accounting for daughter migration in absorbed dose estimates. However, most of this knowledge derives from in vivo models, and there remains a need for complementary in vitro methods that allow for direct, real-time assessment of radiopharmaceutical behavior and retention at the cellular level. 

Conventional measurement techniques often rely on indirect gamma detection \cite{hooijman2024implementing}, which is problematic for isotopes like $^{225}$Ac that emit alpha particles without easily detectable gamma rays. This limitation prevents real-time monitoring of the behavior of alpha emitters in biological systems. Furthermore, conventional alpha detectors, while effective in industrial and scientific settings, are not well-suited for use in biological studies. These detectors typically operate under a vacuum to minimize attenuation, making them incompatible with measurements in living cell cultures. Addressing these challenges requires innovative detection methods that can provide direct and real-time insights into alpha-emitting isotopes under biologically relevant conditions.

This study aims to develop a Bio-Sample Alpha Detector (BAD) and novel measurement protocols for in vitro investigation of \textsuperscript{225}Ac-labeled pharmaceuticals. As a spin-off of a research project related to medical isotopes, we leveraged expertise and technology from the characterization of radioactive ion beams to develop the BAD for direct alpha spectroscopy on cell cultures. By focusing on the development of a silicon PIN (Si-PIN) sensor-based alpha detector capable of operating at ambient temperature and normal atmospheric conditions, we investigated the alpha spectra of pancreatic cancerous cells incubated with labeled \textsuperscript{225}Ac. While this approach provides valuable insights into the behavior of alpha-emitting radiopharmaceuticals at the cellular level, a full understanding of the redistribution and biological impact of recoiling daughter radionuclides will require further evaluations in more complex biological systems, such as animal models. Nonetheless, this study represents an important first step in advancing detection methods for TAT and reducing reliance on animal studies for initial evaluations of radionuclide behavior.

Section 2 outlines experimental configurations and procedures, including a Geant4 simulation of the detector response. Geant4, short for Geometry and Tracking, is a Monte Carlo-based toolkit developed by CERN for simulating the passage of particles through matter, widely used in high-energy physics, medical physics, and space science. In Section 3, experimental data is analyzed, and results are discussed. Section 4 includes a summary and an outlook towards further developments and applications.

\section{Experimental}
\subsection{Material and instruments}

Unless otherwise specified, all solvents and reagents were obtained from commercial suppliers. The F12K Nutrient Mixture (1X), Dulbecco's Modified Eagle Medium (DMEM (1X)), Dulbecco’s Phosphate Buffered Saline (DPBS (1X)), Fetal Bovine Serum (FBS), Penicillin-Streptomycin (PS), and Trypsin-EDTA (0.25$\%$) were purchased from Gibco (USA). The AR42J pancreatic tumor cell line was obtained from BC Cancer Research Centre  (Vancouver, Canada) and was originally sourced from ATCC. Cells were used without additional mycoplasma testing in our laboratory. The crown-TATE targeting agent was synthesized in-house at TRIUMF by Wharton et al \cite{wharton2023preclinical}. The PSMA-617 targeting agent was purchased from MedChemExpress. Ammonium acetate (NH$_4$OAc, trace metal grade), poly-D-lysine, and EDTA were supplied by Sigma-Aldrich. Standard 24-well plates and 150 mm plates, used for routine cell cultures, were purchased from Thermo Fisher Scientific. 
Radiolabeling studies were analyzed using instant thin-layer chromatography (iTLC) with salicylic acid (SA) impregnated iTLC plates (Agilent Technologies). The plates were imaged with an AR2000 TLC scanner (Eckert \& Ziegler) equipped with P10 gas and WinScan V3\_14 software. Activities of $^{225}$Ac were quantified using a high-purity germanium (HPGe) detector (Mirion Technologies, Canberra Inc.), which was calibrated with standardized Europium-152 ($^{152}$Eu) and Barium-133 ($^{133}$Ba) sources in a fixed geometry (20 mL scintillation vials) at specified sample heights \cite{robertson2020232th, yang2020synthesis}. Samples containing $^{225}$Ac were diluted into 20 mL scintillation vials (in H$_2$O), and gamma emission spectra were recorded after secular equilibrium with daughter radionuclides had been reached. All spectra were acquired with a dead time of less than 1.5\%. Radionuclide activities were calculated using Genie 2000 software, which utilized a radioisotope library containing key gamma emission lines of $^{221}$Fr (218 keV) and $^{213}$Bi (440 keV). A Countess 3 automated cell counter, along with Countess cell counting chamber slides (Thermo Fisher Scientific), was used to count cells throughout the experiment. A standard inverted light microscope equipped with a camera was employed to monitor cell cultures, while centrifuges capable of spinning 15 mL and 50 mL Falcon tubes, as well as Eppendorf tubes, were used to create cell pellets. All cell cultures were conducted within a certified level 2 biological safety cabinet (BSC). An Isotemp CO$_2$ incubator maintained at 37 \degree C with 5\% CO$_2$ was used for incubating all cell cultures.

All experimental work was performed in the radiochemistry and cell biology laboratories of TRIUMF's Life Sciences Division. 

\subsection{Bio-sample Alpha Detector}

\begin{figure}[H]
    \centering
    \begin{subfigure}[t]{0.49\textwidth}
        \centering
        \begin{tikzpicture}
                  \node[anchor=south west,inner sep=0] (imageA) at (0,0) 
        {\includegraphics[height=5cm, keepaspectratio]{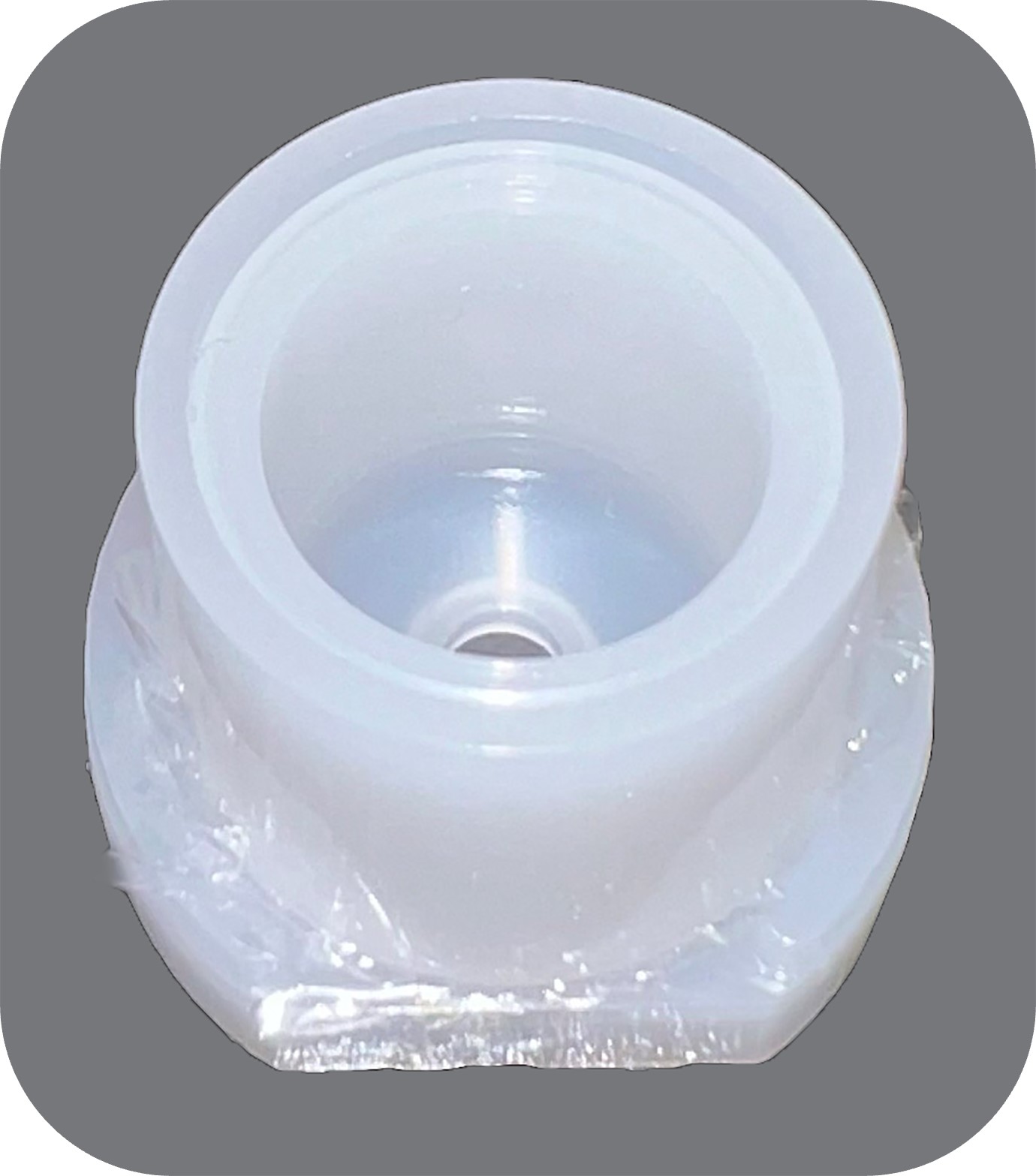}};
        
        \begin{scope}[x={(imageA.south east)},y={(imageA.north west)}]

            \draw[black, thick, <->] (0.75, 0.65) -- (0.25, 0.65)
                node[midway, below, yshift=-3pt, font=\small\sffamily] {\textit{12.1 mm}};

                \draw[black, thick, ->] (1.10, 0.30) node[above, black, xshift=20pt, yshift = -18pt, font=\small\sffamily] {\textit{Mylar Foil}} node[below, xshift=28pt, yshift=-15pt, font=\small\sffamily] {\textit{2.5 $\mu$m thickness}}-- (0.60, 0.53) ; 
                    
            \end{scope}
        \end{tikzpicture}
        \phantomcaption  
        \label{fig:left}  
    \end{subfigure}
    \hspace{0.5em}
    \begin{subfigure}[t]{0.48\textwidth}
        \centering
        \begin{tikzpicture}
            \node[anchor=south west,inner sep=0] (imageB) at (0,0) 
            {\includegraphics[height=5cm, keepaspectratio]{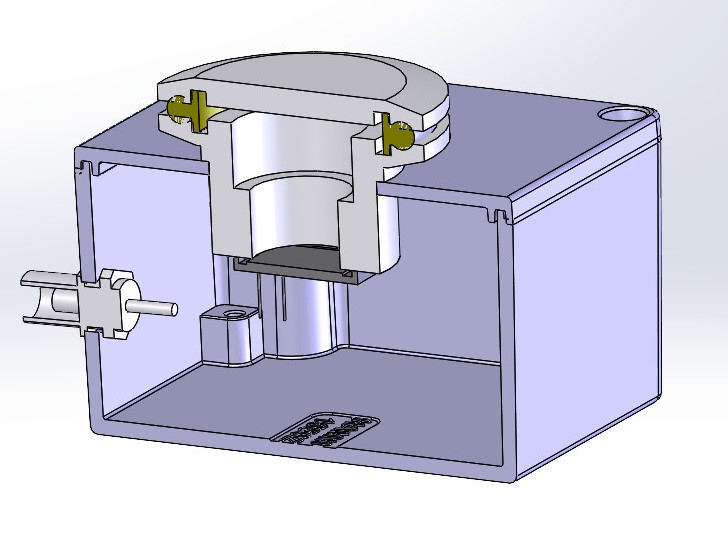}};
            
            \begin{scope}[x={(imageB.south east)},y={(imageB.north west)}]
               
                \draw[black, thick, ->] (0.55, 0.99) node[above, black, font=\small\sffamily] {\textit{Detector}}--  (0.40, 0.49)
                    ;
                \draw[black, thick, ->] (0.15, 0.95)  node[above, black, font=\small\sffamily] {\textit{Sample position}}-- (0.40, 0.60)
                    ;
            \end{scope}
        \end{tikzpicture}
        \phantomcaption 
        \label{fig:right}  
    \end{subfigure}

    \caption{(\textbf{a}): SpectroMicro XRF sample cup with a Mylar foil layer at the bottom, used for cell sample preparation in direct alpha spectroscopy. The cup inner diameter is 12.1 mm, with a Mylar foil thickness of 2.5 $\mu$m to ensure optimal alpha particle detection. (\textbf{b}): CAD 3D cross section view of the detector assembly.}
   
\end{figure}
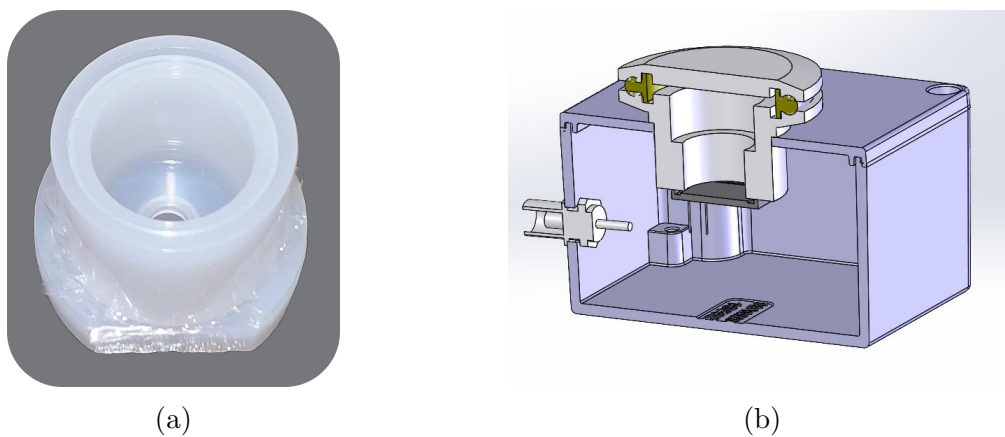

 The BAD utilizes a HAMAMATSU S-series Si-PIN photodiode with an active area of 18 × 18 mm \cite{hamamatsu_s3204}. This photodiode was connected to a Mesytec MSI-8 preamp-shaping amplifier module. A stable bias voltage of +70 V was supplied to the detector using a high voltage power supply from Mesytec. The processed signal from the photodiode was then sent to a 12-bit multi-channel analyzer (Toivel ADC/MCA). Data acquisition was managed with a custom software package using CERN’s ROOT data analysis framework and MIDAS that was adapted for this specific application \cite{kunz2014nuclear}.
For optimal performance, the design of the sample cups was critical. We used SpectroMicro XRF sample cups with dimensions of  18.4 mm (outer diameter), 12.1 mm (inner diameter), and 19.4 mm (height), equipped with a 2.5 \micro m thick Mylar foil layer (~\autoref{fig:left}). The choice of Mylar foil was crucial, as its thinness allows alpha particles to pass through with minimal energy loss, ensuring accurate detection. Additionally, the detector setup was specifically designed to minimize the distance between the radioactive source and the detector, achieving a source-to-detector separation of approximately 100 \micro m. 
The detection system was then calibrated for energy and efficiency using a $^{225}$Ac calibration source prepared by drying a $^{225}$Ac solution of known activity in SpectroMicro XRF sample cups. A Computer-Aided Design (CAD) schematic view of the detector assembly is shown in ~\autoref{fig:right}. The activity in the $^{225}$Ac calibration source was 4 kBq at the time of measurement. The reported activity accounts for both $^{225}$Ac and its decay daughters, as the measurement was performed after secular equilibrium had been reached.

\subsection{$^{225}$Ac production}

The ISAC (Isotope Separation and Acceleration) facility at TRIUMF has recently commenced supplying radionuclides for pre-clinical nuclear medicine research. By irradiating ISOL (Isotope Separation OnLine) targets with a 480 MeV proton beam from the TRIUMF H$^-$ cyclotron, the facility is capable of producing a diverse range of radioactive isotope beams (RIBs). For the production of $^{225}$Ac, composite ceramic uranium carbide targets are employed. These targets typically have thicknesses ranging from 0.05 to 0.1 mol U/cm$^2$ and are irradiated with a proton beam at a maximum current of up to 20 \micro A at an operating temperature of approximately 1950 \degree C. 

A mass-separated RIB of $^{225}$Ac/$^{225}$Ra isotopes is collected at the ISAC Implantation Station, as detailed in \cite{kunz2020medical}. The use of electromagnetic mass separation ensures the isolation of $^{225}$Ac with high isotopic purity, effectively excluding $^{227}$Ac contamination. The $^{225}$Ac is then used either directly or extracted from a $^{225}$Ra generator for experiments in this study \cite{ramogida2019evaluation}. In addition to ISAC, $^{225}$Ac was also sourced from the TRIUMF 520 MeV Isotope Production Facility (IPF). Here, $^{225}$Ac was produced by irradiating $^{232}$Th foils with a $\sim$ 480 MeV proton beam, with a total irradiation dose of up to 12,500 \micro Ah \cite{robertson2019design,robertson2020232th}, However, the $^{225}$Ac used in this study was obtained from a second-pass separation via a $^{225}$Ra generator, which yields $^{225}$Ac with negligible or no $^{227}$Ac contamination, as demonstrated in \cite{robertson2020232th}. 

\subsection{Radiolabeling of $^{225}$Ac-labeled crown-TATE and PSMA-617}

The radiolabeling procedures for both crown-TATE and PSMA-617 with $^{225}$Ac followed similar protocols, with the primary difference being the reaction temperatures used for each compound. crown-TATE (10$^{-4}$ M, 2.5 \micro L) or PSMA-617 (10$^{-4}$ M, 2.5 \micro L) was combined with [$^{225}$Ac]Ac$^{3+}$ (18 kBq/\micro L, 3.5 \micro L) in NH$_4$OAc buffer (pH 7, 5 \micro L) and distilled water (9 \micro L) at room temperature. For crown-TATE, the reaction mixture was gently shaken and incubated at 37 \degree C for 15 minutes, whereas for PSMA-617, incubation occurred at 90 \degree C for 15 minutes \cite{ingham2024preclinical,thakral2021house}. To analyze the radiolabeled products, a 2 \micro L aliquot of the resulting $^{225}$Ac-labeled compound was spotted onto silicic acid-impregnated iTLC plates and developed using EDTA solution (50 mM, pH 5.5). Under these conditions, free [$^{225}$Ac]Ac$^{3+}$ migrated with the solvent front (R$_f$ = 1.0), while both [$^{225}$Ac]Ac-crown-TATE and [$^{225}$Ac]Ac-PSMA-617 remained near the baseline (R$_f$ = 0.0–0.1).  
The TLC plates were imaged with an AR2000 TLC scanner (Eckert \& Ziegler) using P10 gas and WinScan V3\_14 software. To ensure accurate determination of radiochemical yields (RCYs), iTLC imaging of the $^{225}$Ac-labeled products was performed after a 4-hour delay, allowing for the decay of free $^{221}$Fr ($t_{1/2} = 4.80$ min) and $^{213}$Bi ($t_{1/2} = 45.6$ min). 

The molar activity of the radiolabeled compounds was determined using the ratio of the activity of $^{225}$Ac to the number of moles of targeting agent used for radiolabeling. In this study, a molar activity of 256 kBq/nmol was achieved for both crown-TATE and PSMA-617 and used for subsequent experiments. This corresponds to a specific activity of approximately 0.168~kBq/µg for crown-TATE (molecular weight: 1522~g/mol) and 0.202~kBq/µg for PSMA-617 (molecular weight: 1267~g/mol). The ligand-to-metal ratio of 1903:1 indicates that, on average, one $^{225}$Ac atom was coordinated by approximately 1900 targeting peptides in both cases during the radiolabeling process.

$^{225}$Ac-labeled compounds were diluted in DPBS solution to reach the desired volume for the experiment. The diluted activities were further quantified via gamma spectroscopy using a high-purity germanium (HPGe) detector. The activity levels used for each sample type ranged from 3 to 10 kBq. A summary of the radiolabeling conditions and properties for the $^{225}$Ac-labeled compounds is presented in \autoref{tab:labeling-summary}.

\begin{table}[H]
\centering
\footnotesize
\caption{Radiolabeling conditions for each compound used in cell incubation studies.}
\label{tab:labeling-summary}
\begin{tabular}{|l|c|c|c|c|c|}
\hline
\textbf{Compound} & \textbf{Temp. (\textdegree{}C)} & 
\textbf{Time (min)} & \textbf{RCY (\%)} & \textbf{Molar Activity (kBq/nmol)} \\
\hline
\([^{225}\text{Ac}]\text{Ac-crown-TATE}\) & 37 & 15 & 99 & 256 \\
\hline
\([^{225}\text{Ac}]\text{Ac-PSMA-617}\)    & 90 & 15 & 99 & 256 \\

\hline
\end{tabular}
\end{table}

\subsection{ Cell sample preparation}
\subsubsection{Culturing and maintenance of AR42J cells}

The AR42J cell line, derived from a rat pancreatic tumor \cite{atcc_ar42j}, was cultured using standard adherent cell culture techniques. Cells were maintained in a 60:40 mixture of F-12K and  DMEM media, supplemented with 10\% FBS and 1\% PS, at 37 \textdegree C in a humidified incubator with 5\% CO\textsubscript{2}. 
Cells were seeded onto 60 mm tissue culture plates and passaged weekly upon reaching approximately 80\% confluency. The number of passages was kept below 20. For passaging, the medium was aspirated, and the cells were washed with DPBS to remove residual serum. Cells were then detached by adding 2 mL of trypsin-EDTA solution and incubated for 5 minutes at 37 \textdegree C. Trypsin was neutralized with 8 mL of complete media, and the cell suspension was transferred to a Falcon tube for centrifugation at 1300 RPM for 6 minutes. The resulting pellet was resuspended in fresh media and reseeded at a 1:2 dilution, depending on confluency.
AR42J cells are strongly adherent and express a high density of somatostatin receptor 2 (SSTR2), allowing for specific binding of crown-TATE targeting peptides.

\subsubsection{Preparation of cell samples for direct alpha spectroscopy}

 \autoref{tab:cell-prep-workflow} shows the Workflow for the preparation and measurement of AR42J cell samples for alpha spectroscopy. The supernatant media was removed, and the cells were washed with 10 mL of DPBS. Then, the cells were detached from the plate by adding 2 mL of trypsin and incubating at 37 \textdegree C for 5 minutes. The trypsin was then neutralized using 8 mL of media, the cell suspension was transferred to a Falcon tube, and the Falcon tube was taken to the centrifuge. The cells were centrifuged at 1300 RPM (corresponding to approximately 317 g) for 6 minutes. After centrifugation, the media/trypsin supernatant solution was removed, and the cells were resuspended in 5 mL of fresh media. An aliquot of the cell suspension was taken, and the cells were counted using the Countess FL 3. Based on the cell concentration obtained, 300,000 cells were seeded in each well of a 24-well plate. The cells were incubated at 37 \textdegree C for two days to allow sufficient time for cell adhesion and monolayer formation prior to further experimental procedures. After two days, the number of cells was found to be \(4.19 \times 10^5\), with a viability of 85\%.

On the day of the experiment, the media (F12K and DMEM) in each well was carefully removed without disturbing the adherent cell monolayer. 400 $\mu$L of the labeled activity, which was diluted in DPBS, was added to each well, and the cells were incubated at 37 \textdegree C for one hour. Following incubation, the supernatant containing unbound radiolabel was removed, and 200 $\mu$L of trypsin was added to each well to detach the cells from the surface. The well plate was kept in the incubator for 5 minutes to facilitate cell detachment. To neutralize the trypsin, 800 $\mu$L of fresh media was added to each well, bringing the total volume to 1 mL. The cell suspension was then transferred to an Eppendorf tube and centrifuged at 10 G for 6 minutes. After centrifugation, the supernatant was collected in another Eppendorf tube for further analysis, and 400 $\mu$L of fresh media was added to resuspend the cell pellet. The final cell suspension was thoroughly mixed and transferred to SpectroMicro XRF sample cups, which had been pre-coated with poly-D-lysine to promote cell adhesion. It is worth mentioning that simpler methods, such as growing cells directly on Mylar foil, proved inefficient due to \textsuperscript{225}Ac's adherence to Mylar foil, which motivated the development of this refined protocol.

\begin{table}[H]
\centering
\caption{Workflow for the preparation and measurement of AR42J cell samples for alpha spectroscopy.}
\label{tab:cell-prep-workflow}
\resizebox{0.5\textwidth}{!}{
\begin{tabular}{|p{3.5cm}|p{10cm}|}
\hline
\textbf{Step} & \textbf{Procedure} \\
\hline
\textbf{Two days prior to experiment} & 
\begin{itemize}
    \item Prepare sample cups with Mylar foil and coat with poly-D-lysine
    \item Seed cells to 24-well plate
    \item Incubate at 37\,\textdegree{}C for two days
\end{itemize} \\
\hline
\textbf{Cell sample preparation} & 
\begin{itemize}
    \item Remove and discard media above the cell monolayer
    \item Add 400\,µL of labeled activity diluted in DPBS
    \item Incubate cells at 37\,\textdegree{}C for 1 hour
    \item Remove and discard supernatant
    \item Add 200\,µL of trypsin
    \item Incubate for 5 minutes to facilitate cell detachment
    \item Add 800\,µL of fresh media to neutralize trypsin and bring volume to 1\,mL
    \item Transfer suspension to Eppendorf tube and centrifuge at 10\,G for 6 minutes
    \item Collect supernatant for further measurement and analysis
    \item Add 400\,µL of fresh media to pellet and mix thoroughly
    \item Transfer final suspension to sample cup
\end{itemize} \\
\hline
\textbf{Alpha spectrum measurement} & 
\begin{itemize}
    \item Measure cell sample for 900 seconds
\end{itemize} \\
\hline
\end{tabular}
}
\end{table}

\subsubsection{Direct alpha spectroscopy of sell cultures incubated with labeled activity }

\begin{figure}[H]
    \centering
    \includegraphics[width=0.7\textwidth]{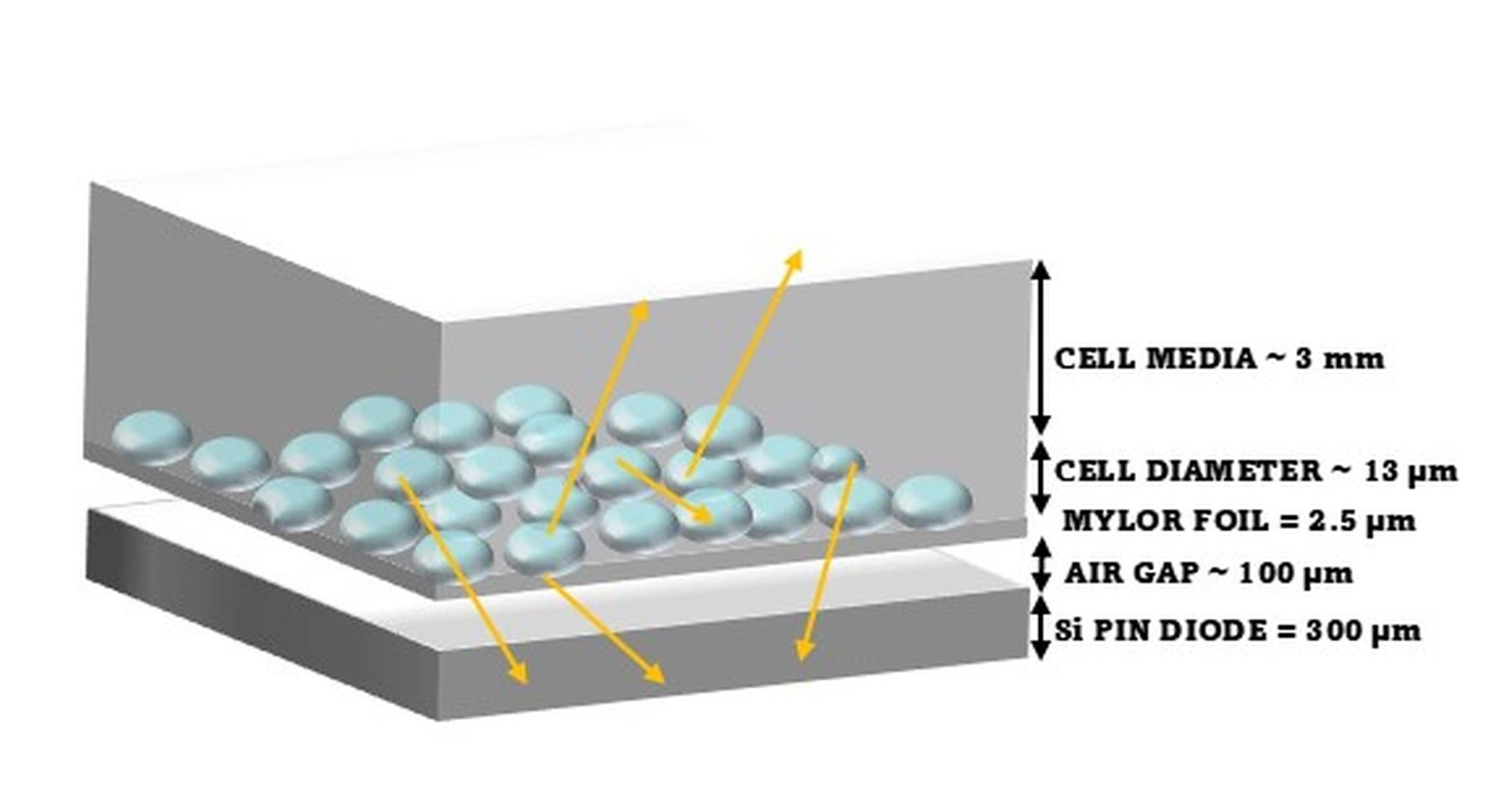} 
    \caption{ Schematic view of the experimental setup for direct alpha spectroscopy, showing the arrangement of AR42J cells with an approximate diameter of 13 \micro m. The cells are placed on a 2.5 µm Mylar foil, with a 100 \micro m air gap separating the cell sample and the Si-PIN diode (300 \micro m thickness). The media layer above the cells is approximately 3 mm thick. The yellow arrows represent the trajectory of emitted alpha particles.}
    \label{fig:experimental setup} 
\end{figure}

Three sets of sample cups were prepared as described in Section 2.5. The first set contained [$^{225}$Ac]Ac-crown-TATE, the second set contained [$^{225}$Ac]Ac-PSMA-617 as a negative control (assuming the AR42J cells do not have receptors for PSMA-617), and the third set was incubated with [$^{225}$Ac]Ac$^{3+}$. For each set, a corresponding series of reference sample cups was prepared with the same amount of labeled activity and following the identical procedure, but without cells. Each sample cup was placed on the detector, and spectra were acquired for 900 seconds, achieving a statistical uncertainty of less than 1\% in the total recorded counts. The experiment was repeated three times, and the standard deviation was applied to the results. \autoref{fig:experimental setup} illustrates a schematic view of the experimental setup.

\subsubsection{Geant4 simulation}

In this study, we employed the Geant4 Monte Carlo simulation toolkit (version 11.2.2) to model the interaction of the radioisotope $^{225}$Ac in an aqueous solution with a silicon photodiode detector. Geant4 is a widely used tool for simulating the passage of particles through matter, allowing for detailed predictions of particle interactions within various materials \cite{agostinelli2003geant4}. To simulate the physics of the system, we used a custom physics list that included several physics processes critical to our study: electromagnetic interactions (via the G4EmStandardPhysics class), optical photon interactions (G4OpticalPhysics), radioactive decay processes (G4RadioactiveDecayPhysics), and general particle decay (G4DecayPhysics). Given the complexity of accurately simulating biological cells at the microscale, we used varying thicknesses of water layers, representing different depths from the bottom of the sample cups, as an alternative for cell layers. This approach enabled a more realistic approximation of the experimental conditions and facilitated comparison between the simulated energy spectra and experimental results.

The silicon photodiode detector used in the simulation had dimensions of 18~$\times$~18~mm and a thickness of 300~$\mu$m. A cylindrical sample cup with a diameter of 13~mm and a height of 3~mm was separated by a 100~$\mu$m air gap from the detector surface. A 2.5~$\mu$m thick Mylar foil was placed at the bottom of the sample cups.
To model the radioactive source, $^{225}$Ac atoms were randomly distributed within a defined cylindrical volume (using the \texttt{G4Tubs} class), either as a thin layer on the Mylar foil or within a 13~$\mu$m-thick water layer representing a cell monolayer. The position of each $^{225}$Ac nucleus was assigned randomly within this volume using uniform random distributions (\texttt{G4UniformRand()} function), approximating a uniform source distribution. Only the dominant alpha emissions from $^{225}$Ac and its primary daughters were simulated; weaker decay branches were not considered. Each simulation was performed with $10^6$ primary events to ensure sufficient statistical accuracy.
The simulated spectra for the Mylar surface and 13~$\mu$m water layer cases are presented in \autoref{fig:sim-exp-nocell} and \autoref{fig:sim-exp-cell}, respectively, alongside experimental data for comparison. The weaker decay branches were not considered in this simulation.

\section{Results and discussion}
\subsection{Alpha decay spectrum from $^{225}$Ac calibration source}

\begin{figure}[H]
    \centering
   
    \begin{tikzpicture}
       
        \node[anchor=south west, inner sep=0] (image) at (0,0) 
        {\includegraphics[width=0.5\textwidth]{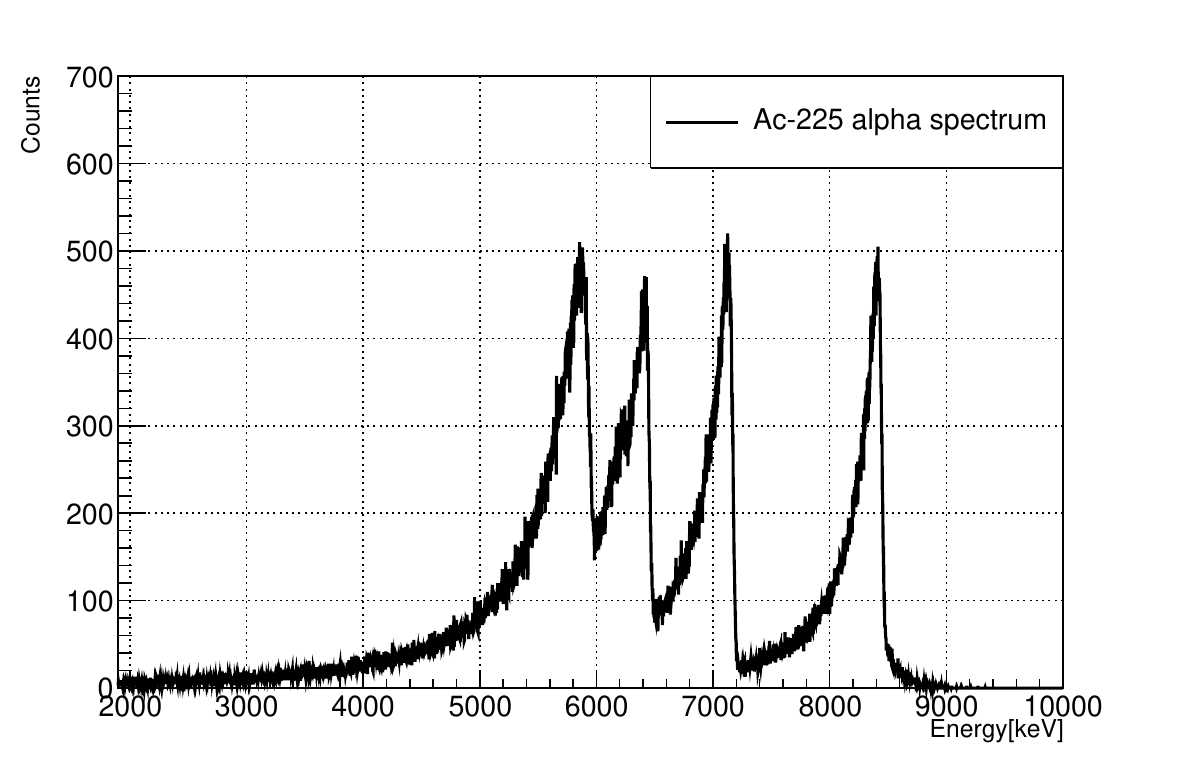}};

        \begin{scope}[x={(image.south east)},y={(image.north west)}]
           
            \node[anchor=west] at (0.35,0.70) {\scriptsize($^{225}$Ac)};
            \node[anchor=west] at (0.48,0.72) {\scriptsize($^{221}$Fr)};
            \node[anchor=west] at (0.61,0.73) {\scriptsize($^{217}$At)};
            \node[anchor=west] at (0.75,0.70) {\scriptsize($^{213}$Po)};
        \end{scope}
    \end{tikzpicture}
    
    \caption{Alpha spectrum of $^{225}$Ac obtained using the Bio-sample Alpha Detector (BAD). The four distinct peaks correspond to the alpha decays of $^{225}$Ac (5830 keV), $^{221}$Fr (6340 keV), $^{217}$At (7067 keV), and $^{213}$Po (8376 keV). The spectrum demonstrates the clear detection of each isotope in the decay chain.}
    \label{fig:Ac-225} 
\end{figure}

\autoref{fig:Ac-225} shows the alpha spectrum for $^{225}$Ac obtained using the BAD detection system, revealing four distinct peaks for $^{225}$Ac (5830 keV), $^{221}$Fr (6340 keV), $^{217}$At (7067 keV), and $^{213}$Po (8376 keV). Calibration was performed using the high-energy flank of each peak. The characteristic tailing toward lower energies is attributed to alpha particle energy loss as they pass through both the 2.5~$\mu$m Mylar foil and the $\sim$100~$\mu$m air gap between the sample and the detector surface. The air gap dimension was determined from design drawings, as it could not be directly measured after assembly. The energy loss in air (12 keV per 100~$\mu$m) is relatively minor compared to the larger energy loss in the Mylar foil (260 keV for 2.5~$\mu$m thickness). Importantly, because the detector was calibrated using a dried $^{225}$Ac source placed on the Mylar within the final sample cup geometry, any energy loss contributions from both the Mylar and air gap are already incorporated into the calibration. Thus, small deviations between the design and physical assembly would not significantly affect the measurement accuracy. The overall detector efficiency across the entire energy range was calculated to be 37\%.

\subsection{Cell culture distribution}

\begin{figure}[H]
    \centering
    \includegraphics[width=0.5\textwidth]{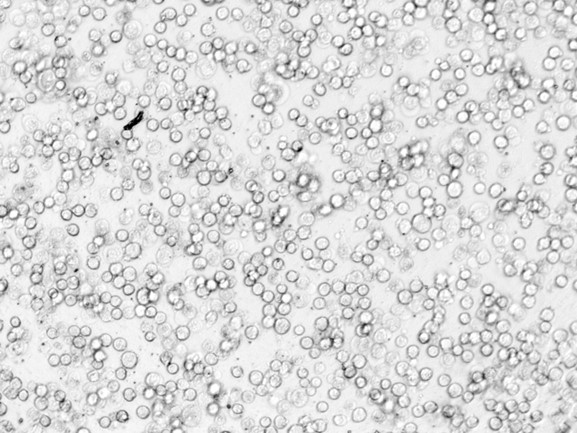} 
    \caption{ Microscopic image of AR42J rat pancreatic tumor cells in culture. The cells are shown adhering to the surface, displaying their characteristic round morphology. This image was taken before incubation with radiolabeled $^{225}$Ac compounds for direct alpha spectroscopy.}
    \label{fig:cell-distribution} 
\end{figure}

\begin{figure}[H]
    \centering
    \begin{subfigure}[t]{0.5\textwidth}
        \centering
        \begin{tikzpicture}
            \node[anchor=south west,inner sep=0] (main) at (0,0) 
            {\includegraphics[height=5cm]{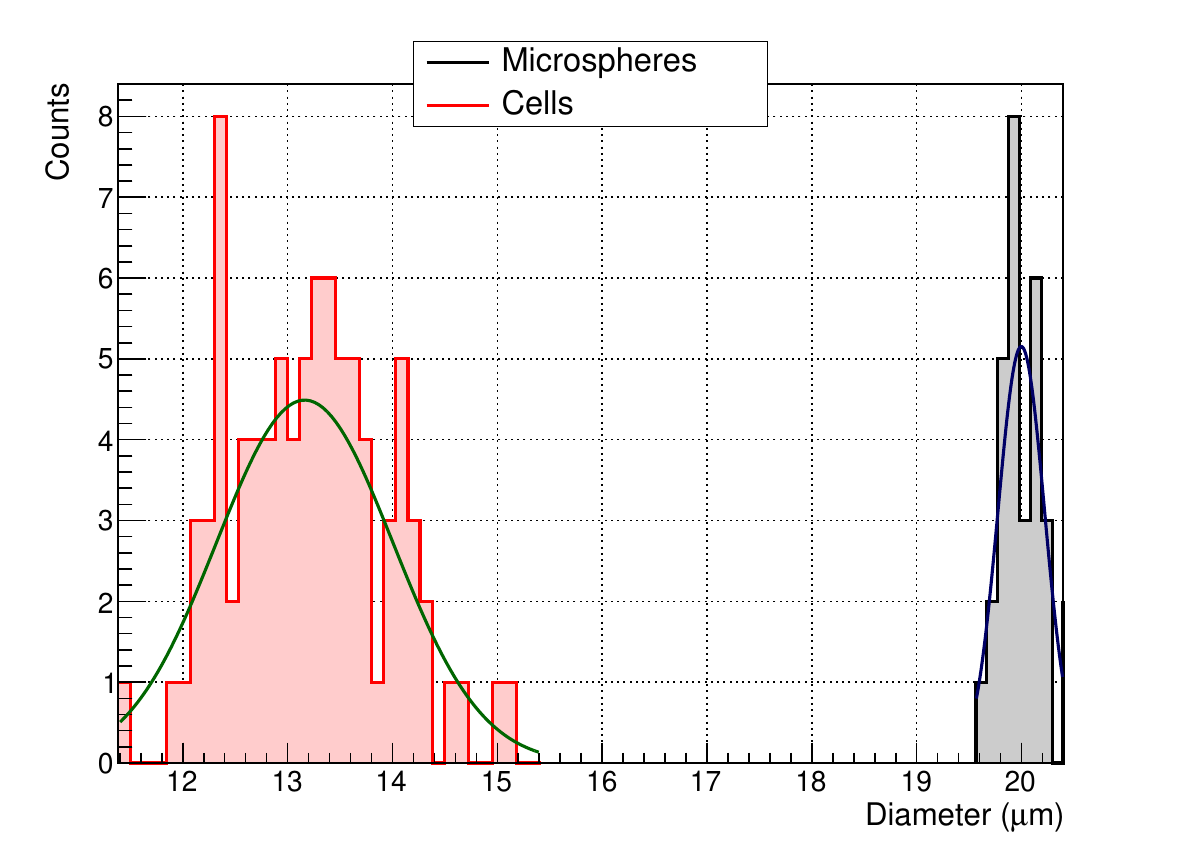}};
            
            \begin{scope}[x={(main.south east)}, y={(main.north west)}]
                \node[anchor=south west] at (0.4, 0.4)
                {\includegraphics[width=0.3\textwidth]{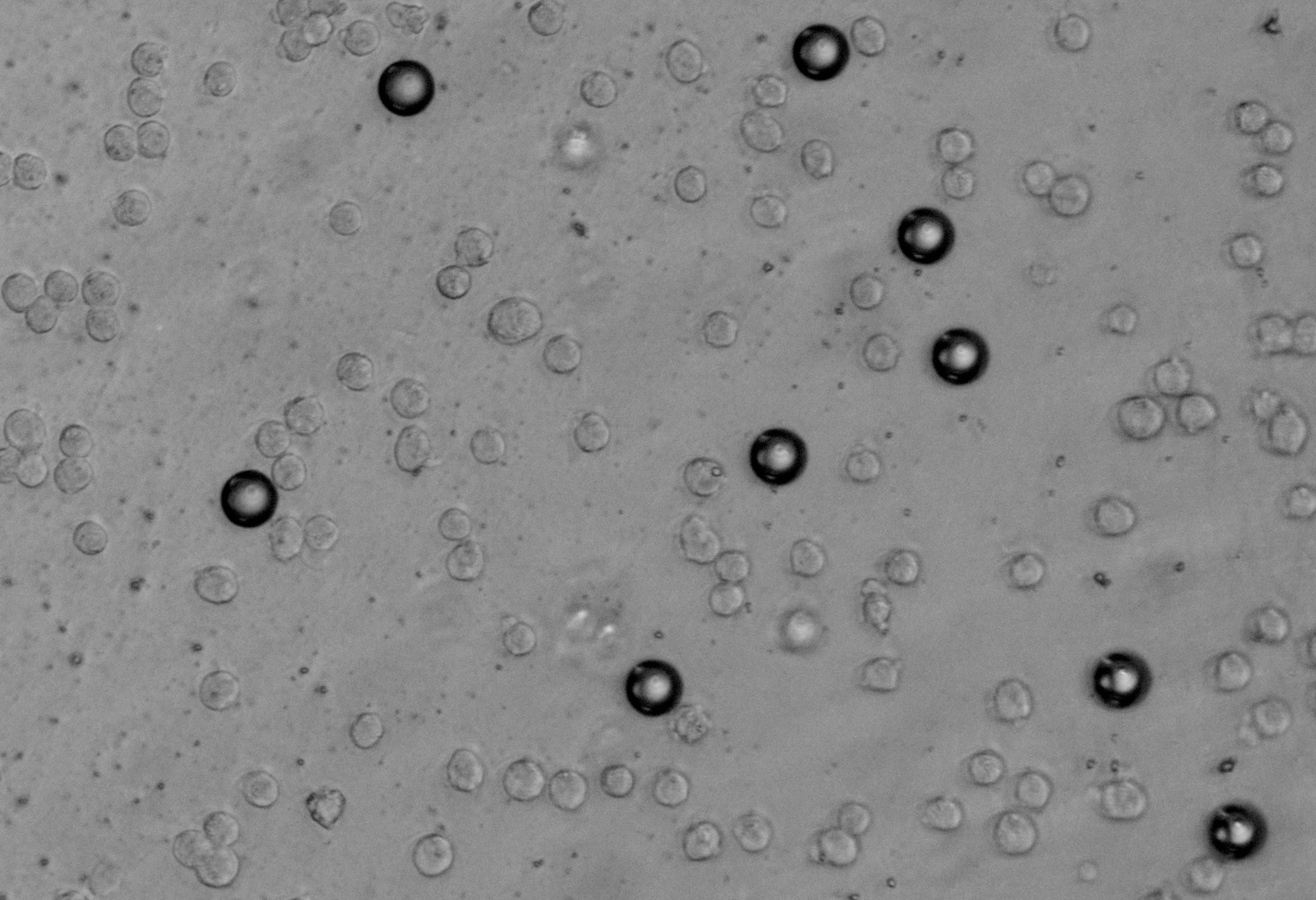}};
            \end{scope}
        \end{tikzpicture}
        \caption{}
        \label{fig:cell-hist}
    \end{subfigure}
    \hspace{-1.5em}
    \begin{subfigure}[t]{0.5\textwidth} 
        \centering
        \includegraphics[height=5cm]{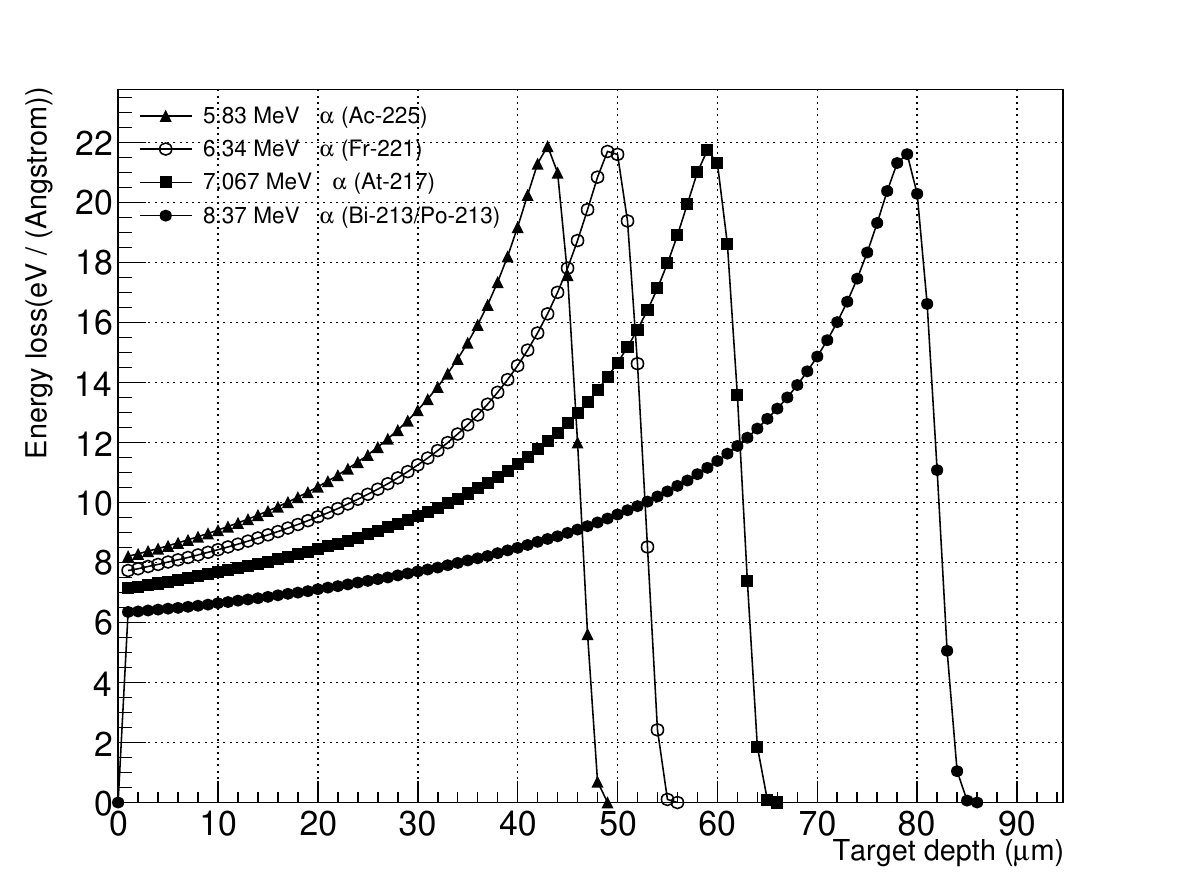} 
        \caption{}
        \label{fig:bragg-peak}
    \end{subfigure}

    \caption{(\textbf{a}) Normalized distribution of microspheres and AR42J cells based on diameter, with an inset showing a microscopic image of cells with reference spheres. Microspheres (black) have a diameter of 20~$\mu$m, while cells (red) show a wider distribution centered around 13.16~$\pm$~0.12~$\mu$m.
    (\textbf{b}) Stopping power curves (Bragg peaks) simulated in SRIM for $\alpha$-particles emitted by $^{225}$Ac and its decay daughters in water. The 5.83~MeV $\alpha$-particles from $^{225}$Ac exhibit a range from approximately 40~$\mu$m to 50~$\mu$m, indicating that particles originating inside AR42J cells ($\sim$13~$\mu$m in diameter) can exit the cells and be detected.}
    \label{fig:size}
\end{figure}

\autoref{fig:cell-distribution} shows the distribution of AR42J cells on the Mylar foil. The cells appear to be evenly distributed across the surface of the sample cup, with no visible clumping or aggregation. This even distribution is crucial for ensuring uniform exposure during alpha spectroscopy. Microspheres with a known diameter of 20.1 $\mu$m were used as a reference, and the cell diameter was determined based on these microspheres using a microscope image and ImageJ software. A Gaussian fit was applied to the distribution, resulting in a cell diameter of 13.16 $\pm$ 0.12 $\mu$m. \autoref{fig:cell-hist} shows the distribution of cell diameters in comparison with the microspheres. The size of the cells is important because it affects the ability of alpha particles to exit the cells and be detected before losing significant energy. An SRIM (Stopping and Range of Ions in Matter) simulation was performed to calculate the range of alpha particles in a water-based material, approximating the composition of the biological cells. The simulation included alpha particles emitted from $^{225}$Ac and its alpha-emitting daughters, using their characteristic energies. \autoref{fig:bragg-peak} shows the Bragg peak from the SRIM simulation, indicating that alpha particles from $^{225}$Ac have a range of approximately 50~$\mu$m in water. Considering that the average AR42J cell diameter is around 13~$\mu$m, the simulation confirms that alpha particles can escape the cells and reach the detector. This capability is essential for directly measuring the uptake and behavior of alpha-emitting radiopharmaceuticals.

\subsection{Alpha spectrum from cell sample incubated with [$^{225}$Ac]Ac-crown-TATE }

\begin{figure}[H]
    \centering
    \begin{subfigure}[t]{0.5\textwidth}
        \centering
        \includegraphics[height=5cm, keepaspectratio]{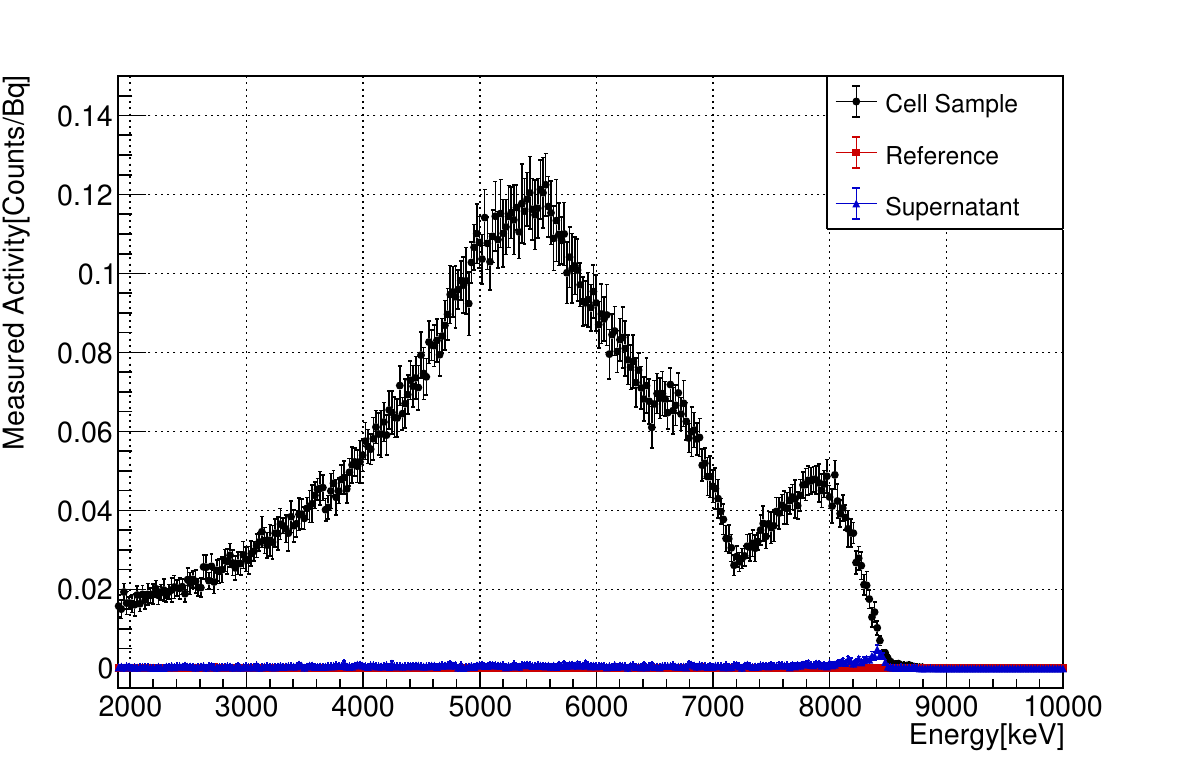} 
        \caption{}
        \label{fig:CT-Ac225}
    \end{subfigure}
    \hspace{-2.em}
    \begin{subfigure}[t]{0.5\textwidth} 
        \centering
        \includegraphics[height=5cm, keepaspectratio]{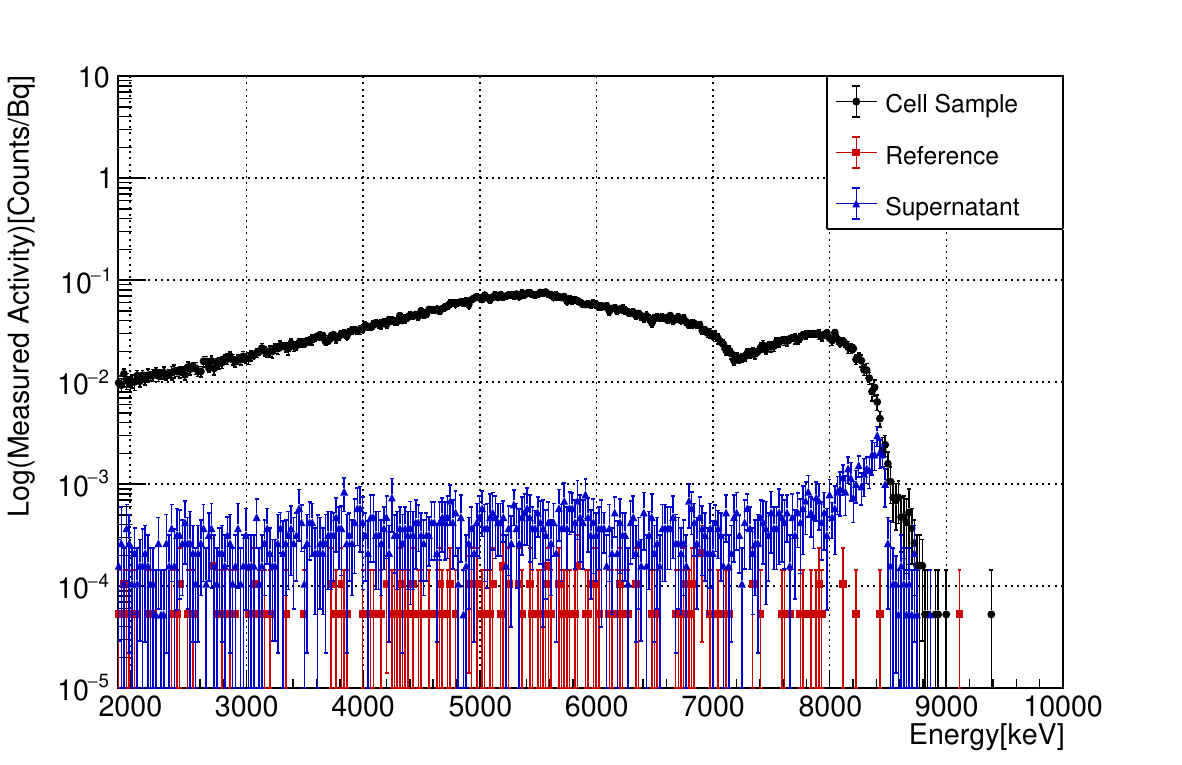} 
        \caption{}
        \label{fig:CT-Ac225-log}
    \end{subfigure}
    \caption{Alpha spectra of [$^{225}$Ac]Ac-crown-TATE obtained from the cell sample (black), supernatant (blue), and reference sample (red, no cells) (\textbf{a}):  Linear scale. (\textbf{b}):  Logarithmic scale.}
    \label{fig:Crown-Tate-Ac225}
\end{figure}

\autoref{fig:CT-Ac225} presents the alpha spectrum of [$^{225}$Ac]Ac-crown-TATE in  the cell sample, alongside the corresponding reference and post-wash supernatant samples, acquired using the BAD detection system. The same spectra are shown on a logarithmic scale in \autoref{fig:CT-Ac225-log} for better visualization of lower-intensity peaks. Although the detector resolution is insufficient to fully resolve closely spaced peaks such as those of $^{225}$Ac and $^{221}$Fr, peaks consistent with  $^{217}$At and $^{213}$Po remain identifiable at their expected energy positions. The number of counts was normalized to the initial activity measured by a calibrated HPGe detector, enabling direct comparison between the three sample types. Three replicates of each sample were prepared, and the standard deviation is reflected in the plot. As observed in the spectrum from the cell sample, the alpha peaks appear broader compared to the dried sample, which is due to energy loss as alpha particles traverse through cellular material in aqueous media before reaching the detector. Interestingly, a slight energy shift of the $^{213}$Po peak is seen in the cell sample compared to the post-wash supernatant. While this shift might suggest increased energy loss in the presence of cells, the exact mechanism behind this observation remains uncertain, and no definitive conclusion can be drawn without further investigation.
The supernatant shown in this figure refers to the media collected after the final wash step, following trypsinization and centrifugation, and thus contains minimal residual unbound activity. This signal corresponds to approximately $\sim$1.17\% of the cell-associated signal across the entire energy range. The $^{213}$Po peak is visible in the supernatant, likely reflecting free $^{213}$Bi released from the radiopharmaceutical complex due to the nuclear recoil of $^{225}$Ac decay. As previously described, such recoil can disrupt the coordination bonds of the labeled compound and cause daughter radionuclides to migrate into the surrounding solution.
The reference sample, which contained no cells and underwent the same radiolabeling and washing procedures, exhibited only minimal detectable counts in the alpha spectrum, corresponding to approximately $\sim$0.09\% of the signal from the cell-associated fraction. This indicates that unbound radiolabeled compounds were effectively removed during sample preparation.

\subsection{Alpha spectrum from cell sample incubated with [$^{225}$Ac]Ac-PSMA-617}

\begin{figure}[H]
    \centering
    \begin{subfigure}[t]{0.5\textwidth}
        \centering
        \includegraphics[height=5cm, keepaspectratio]{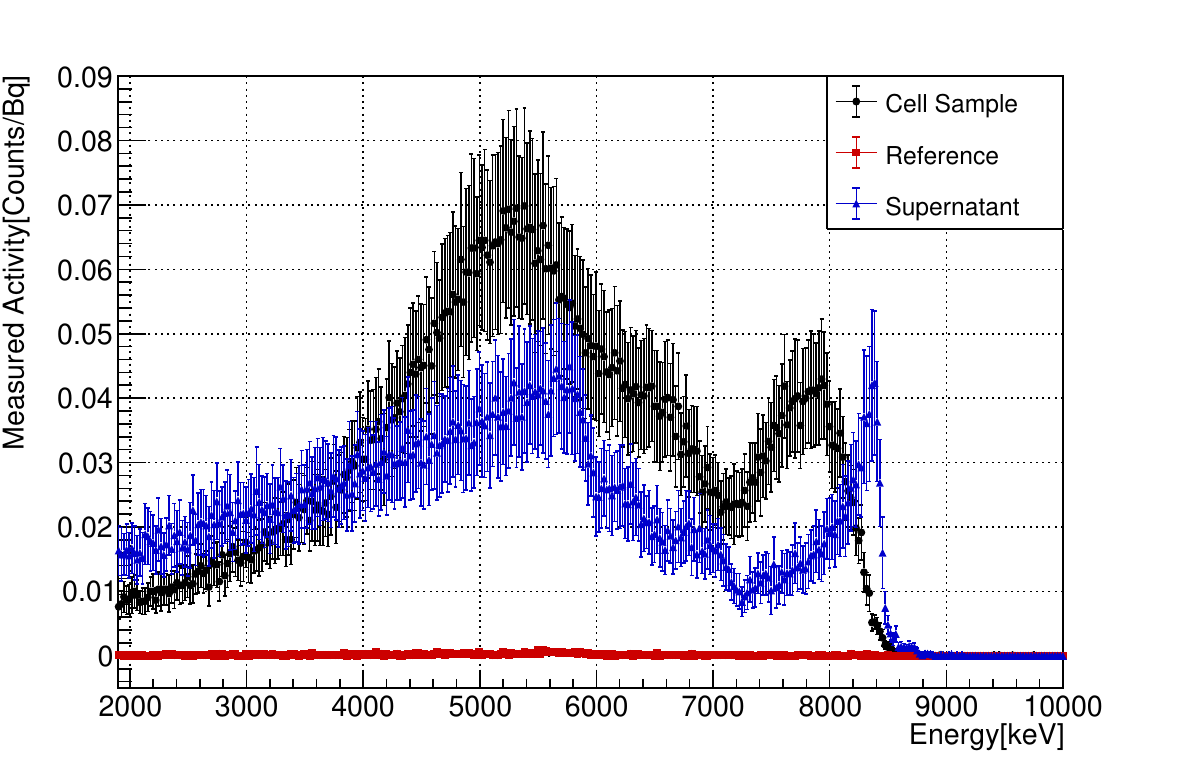} 
        \caption{}
        \label{fig:PSMA-Ac225}
    \end{subfigure}
    \hspace{-2.em}
    \begin{subfigure}[t]{0.5\textwidth} 
        \centering
        \includegraphics[height=5cm, keepaspectratio]{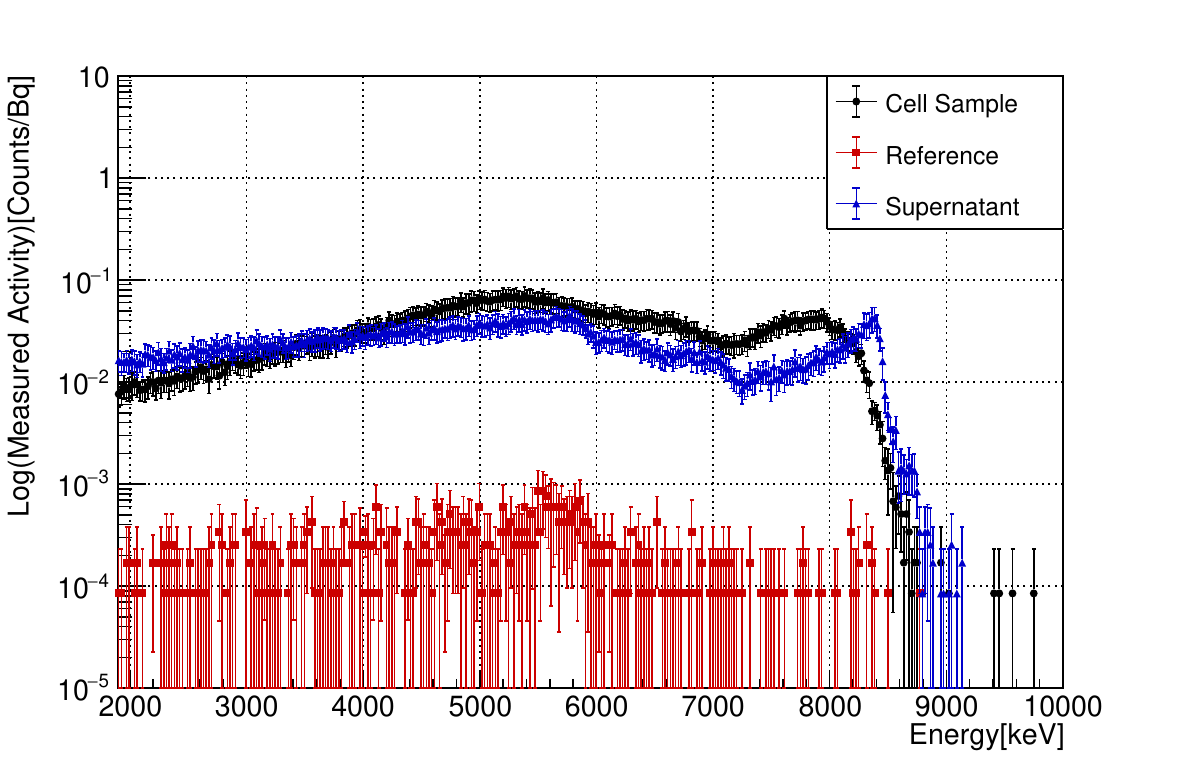} 
        \caption{}
        \label{fig:PSMA-Ac225-log}
    \end{subfigure}
    \caption{Alpha spectra of [$^{225}$Ac]Ac-PSMA-617 obtained from the cell sample (black), reference sample (red), and supernatant (blue). (\textbf{a}):  Linear scale. (\textbf{b}):  Logarithmic scale.}
    \label{fig:PSMA-Ac225-1}
\end{figure}

As a negative control, the behavior of [$^{225}$Ac]Ac-PSMA-617 on AR42J cell cultures was investigated. \autoref{fig:PSMA-Ac225} shows the alpha spectrum of [$^{225}$Ac]Ac-PSMA-617 from the cell sample, along with the reference and cell supernatant samples, acquired using the BAD detection system in linear scale, while \autoref{fig:PSMA-Ac225-log} presents the spectrum on a logarithmic scale. Since AR42J cells lack receptors for the PSMA-617 ligand, minimal uptake of [$^{225}$Ac]Ac-PSMA-617 by the cells was expected. The alpha spectrum from the cell sample is almost identical to that of the supernatant. Comparison of the net area under the spectra across the entire energy range indicates that the supernatant signal corresponds to approximately $\sim$79.5\% of the cell-associated signal, suggesting that[$^{225}$Ac]Ac-PSMA-617 remained largely unbound and was released into the medium following centrifugation. The reference sample, which lacks cells, exhibits very low counts in the spectrum, approximately $\sim$0.55\% of the cell-associated signal, indicating  that the washing procedure effectively removed unbound [$^{225}$Ac]Ac-PSMA-617. Additionally, as mentioned in the previous section, the observed energy shift in the cell sample may be attributed to interactions with the cells and requires further investigation.

\subsection{Alpha spectrum from cell sample incubated with  [$^{225}$Ac]Ac$^{3+}$}

\begin{figure}[H]
    \centering
    \begin{subfigure}[t]{0.5\textwidth}
        \centering
        \includegraphics[height=5cm, keepaspectratio]{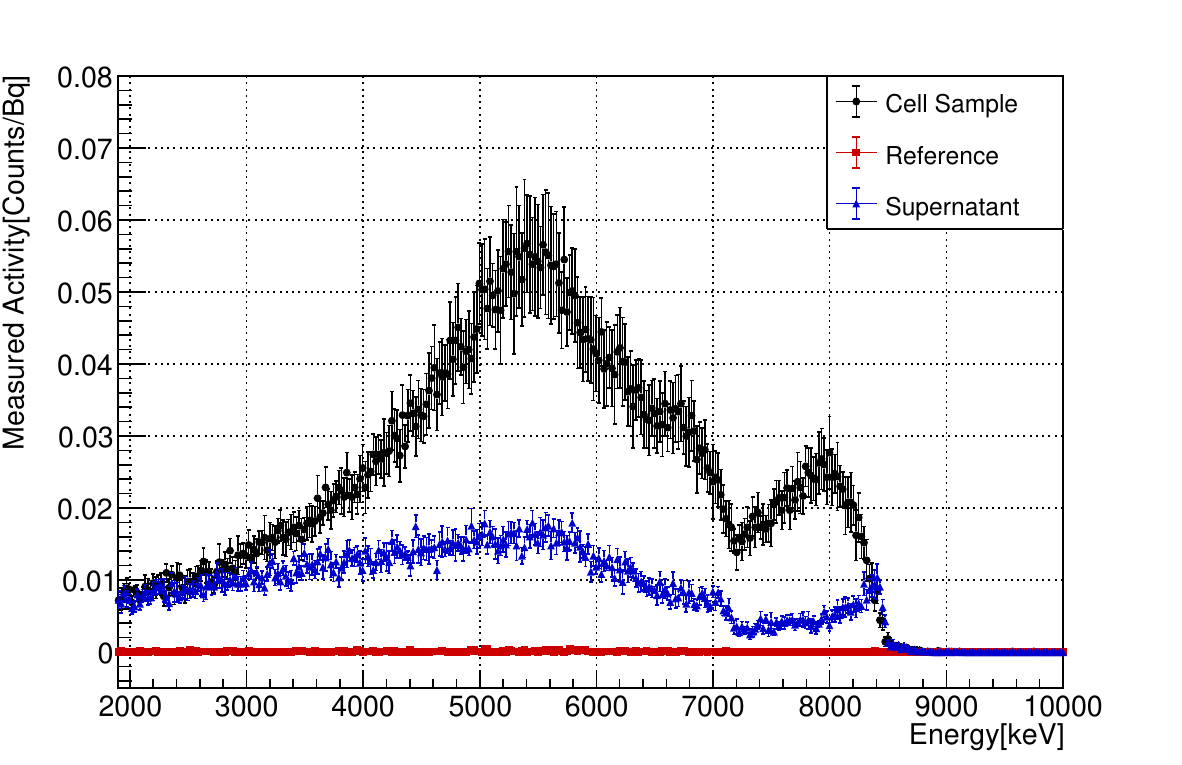} 
        \caption{}
        \label{fig:Free-Ac225}
    \end{subfigure}
    \hspace{-2.em}
    \begin{subfigure}[t]{0.5\textwidth} 
        \centering
        \includegraphics[height=5cm, keepaspectratio]{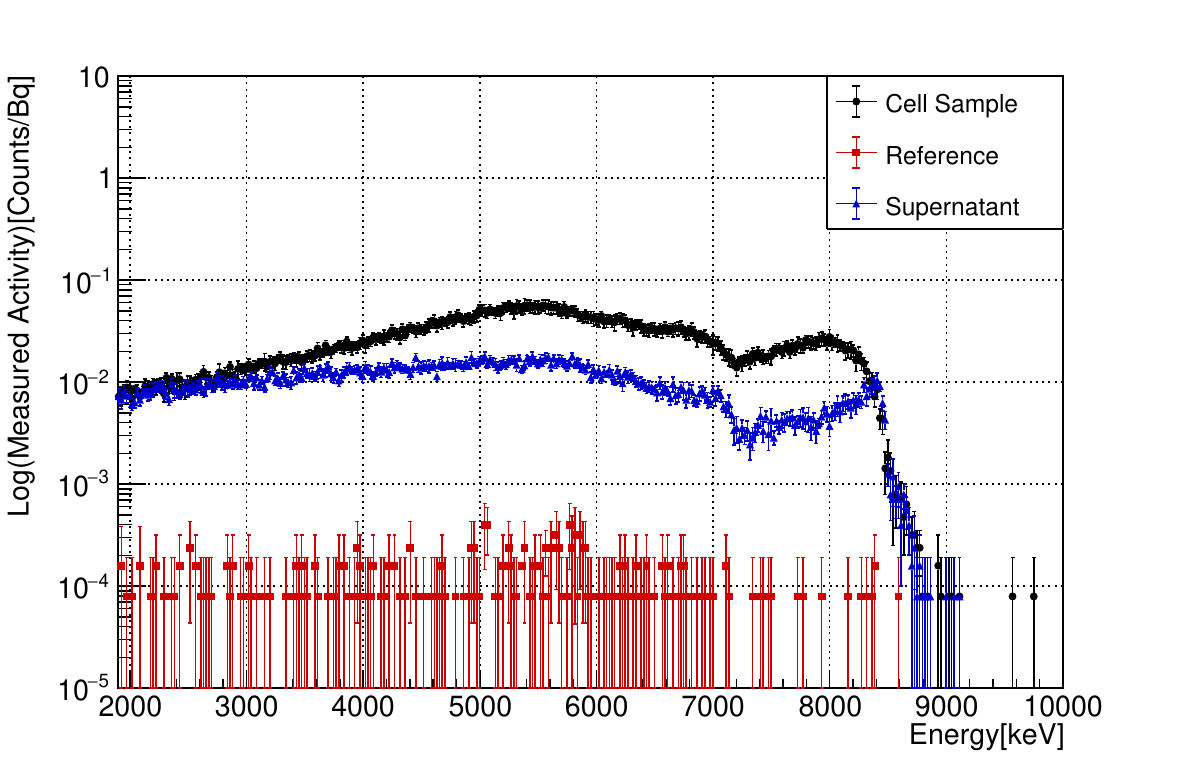} 
        \caption{}
        \label{fig:Free-Ac225-log}
    \end{subfigure}
    \caption{Alpha spectra of free [$^{225}$Ac]Ac$^{3+}$ obtained from the cell sample (black), supernatant (blue), and reference sample (red, no cells) (\textbf{a}):  Linear scale. (\textbf{b}):  Logarithmic scale.}
    \label{fig:Crown-Tate-Ac225}
\end{figure}

The uptake of free  [$^{225}$Ac]Ac$^{3+}$ in AR42J cells  was studied, and the alpha spectra of the cell sample, supernatant, and reference (no cells) were acquired using the BAD detection system, as shown in \autoref{fig:Free-Ac225} and \autoref{fig:Free-Ac225-log}. 
These spectra indicate that free [$^{225}$Ac]Ac$^{3+}$ is taken up by the cells, but to a lesser extent than [$^{225}$Ac]Ac-crown-TATE (approximately 76\% of the total cell-associated signal).
By comparing the spectra of the cell sample and the supernatant, it is evident that free [$^{225}$Ac]Ac$^{3+}$ remains in the supernatant, with the net area under the supernatant spectrum across the entire energy range accounting for 40.1\% of the cell sample signal. This finding suggests partial uptake of the isotope by the cells. As expected, the reference sample, which contains no cells, shows minimal activity (approximately 0.25\% of the cell sample signal) confirming that the washing procedure effectively removes unbound activity.

\subsection{Geant4 simulation}

\begin{figure}[H]
    \centering
    \includegraphics[width=0.5\textwidth]{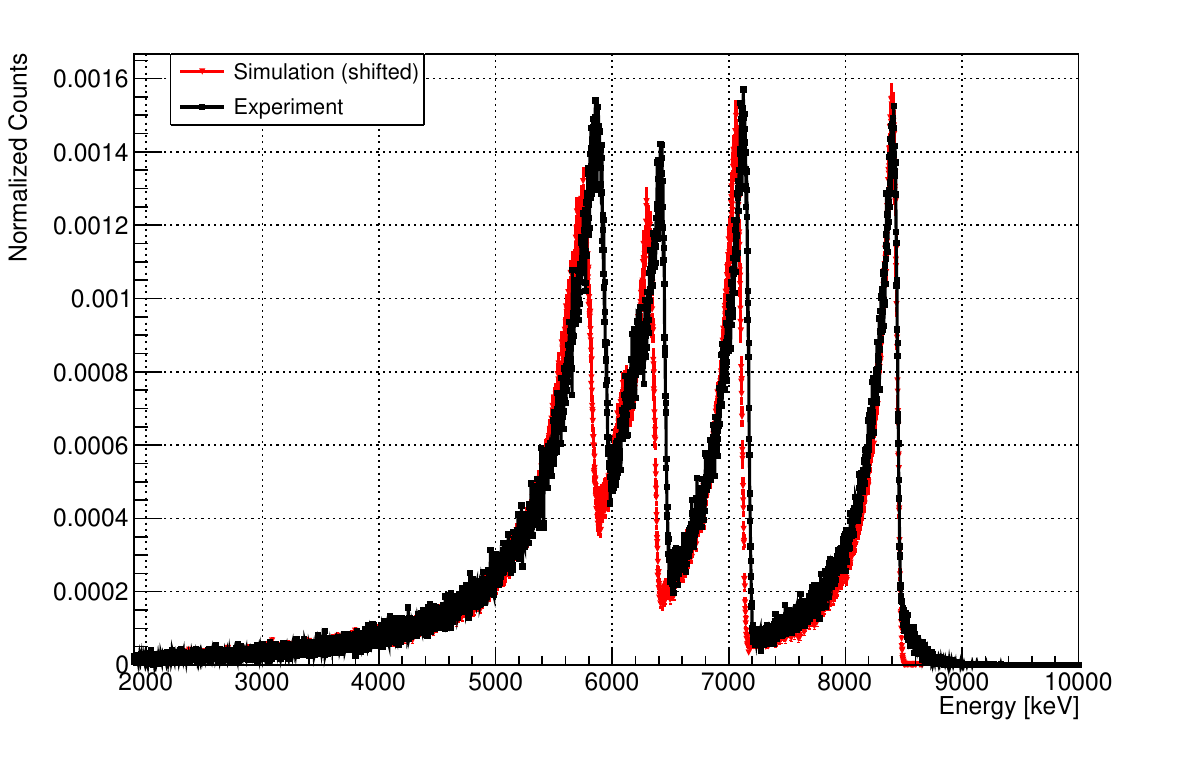} 
    \caption{ Comparison of experimental and simulated alpha spectra for dried $^{225}$Ac placed on the Mylar foil near the detector. }
    \label{fig:sim-exp-nocell} 
\end{figure}

\autoref{fig:sim-exp-nocell} shows a comparison between the experimental and simulated spectra for $^{225}$Ac distributed on the Mylar foil near the detector. The four characteristic peaks of the $^{225}Ac$  decay chain $^{225}Ac$ (5830 keV), $^{221}$Fr (6340 keV), $^{217}$At (7067 keV), and Polonium-213 ($^{213}$Po) (8376 keV) are clearly distinguishable. The simulation and experimental data are well-aligned, with a particularly strong match at the $^{213}$Po peak. The slight discrepancies observed in the other peaks may be attributed to minor differences in the physical setup, such as detector response or material variations in the sample cup that were not fully captured in the simulation. 

\begin{figure}[H]
    \centering
    \includegraphics[width=0.5\textwidth]{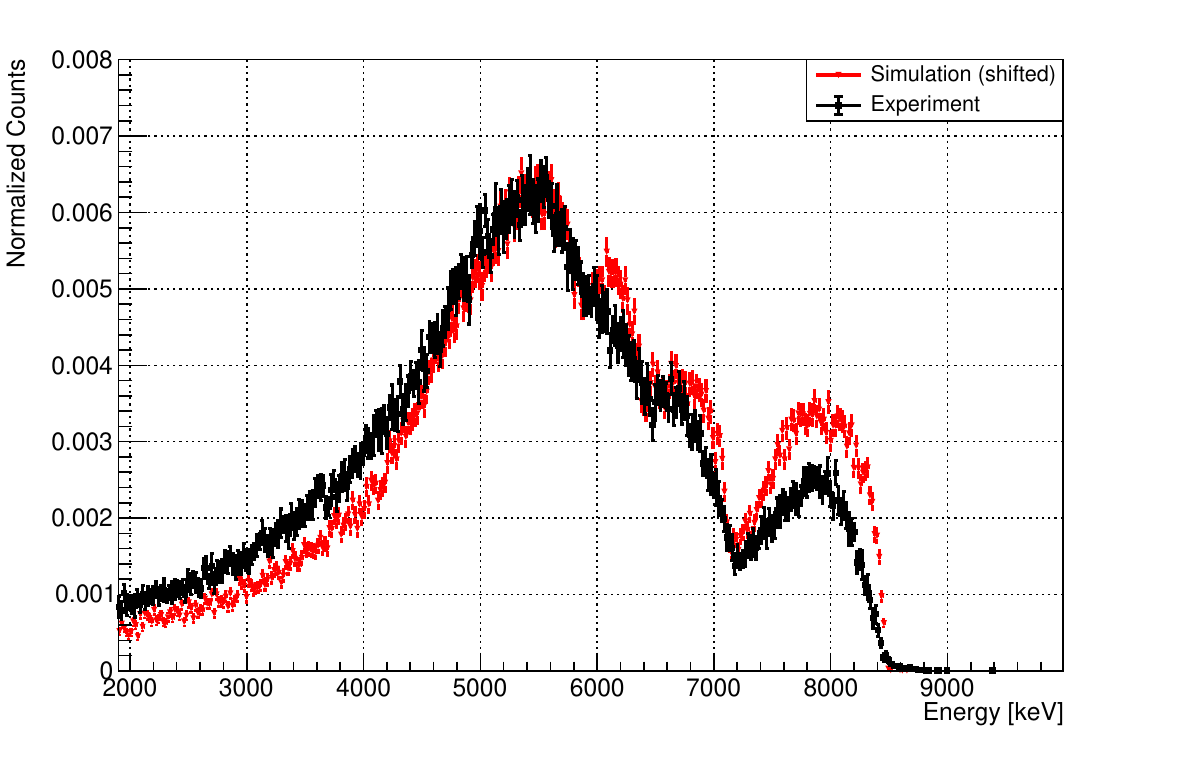} 
    \caption{ Comparison of experimental and simulated alpha spectra for $^{225}$Ac randomly distributed in an aqueous solution, representing cell samples with a thickness of 13 \micro m.  }
    \label{fig:sim-exp-cell} 
\end{figure}

\autoref{fig:sim-exp-cell} compares the experimental and simulated spectra for $^{225}$Ac distributed in the cell medium. In the Geant4 simulation, $^{225}$Ac was modeled as being randomly distributed in an aqueous solution with a thickness of 13 \micro m, representing the approximate diameter of the cells. The experimental and simulated spectra show good agreement, confirming the behavior of $^{225}$Ac in solution. This alignment demonstrates that the simulation accurately models the energy deposition and supports the validity of the experimental approach. It is important to highlight that a 300 keV shift was applied to the simulation data to account for energy loss caused by the Mylar foil, as calculated using the SRIM software. This energy loss is not observed in the experimental data because the detector was calibrated with the Mylar foil in place. This calibration inherently compensates for the energy loss, ensuring accurate energy measurements in the experimental spectrum. 

\section{Conclusion and outlook}

To the best of our knowledge, this study is the first to employ an alpha detector for direct alpha spectroscopy of cancer cell samples under ambient conditions, demonstrating the potential of the Bio-sample Alpha Detector (BAD) as a novel method in the context of targeted alpha therapy. Presented as a proof-of-principle, this study enabled the direct measurement of alpha spectra from AR42J rat pancreatic tumor cells incubated with [\textsuperscript{225}Ac]Ac-crown-TATE, [\textsuperscript{225}Ac]Ac-PSMA-617, and [\textsuperscript{225}Ac]Ac\textsuperscript{3+}. The results were interpreted qualitatively and indicate significant uptake of [\textsuperscript{225}Ac]Ac-crown-TATE by the cells, as evidenced by distinct alpha spectral features and minimal counts in the supernatant, suggesting efficient retention. In contrast, limited uptake of [\textsuperscript{225}Ac]Ac-PSMA-617 was observed, consistent with the absence of PSMA receptors on AR42J cells.

Detection of \textsuperscript{213}Po, a decay product of \textsuperscript{225}Ac, serves as an indicator of \textsuperscript{213}Bi redistribution from the cells into the surrounding media. This redistribution impacts dosimetry and toxicity, as progeny isotopes like \textsuperscript{213}Bi can migrate to off-target tissues, potentially increasing toxicity risks. The well-documented kidney toxicity of \textsuperscript{213}Bi \cite{schwartz2011renal} underscores the importance of understanding and mitigating this redistribution to optimize the therapeutic index of radiopharmaceuticals. Geant4 simulations validated experimental results and provided insights into alpha particle interactions with biological samples.

This paper serves as an initial validation of the BAD system and introduces it as a potential tool for studying alpha-emitting radiopharmaceuticals in vitro.
In future work, we plan to develop a multi-channel BAD array (e.g., a 12-channel detector), integrated with a CO\textsubscript{2} incubator and automated flow system, to enable quantitative, high-throughput studies. These improvements will allow for higher statistical power and more systematic evaluations of uptake, retention, and redistribution under physiologically relevant conditions. A peristaltic pump system to control radiolabeled compound flow through cell cultures will simulate vascular conditions, providing insights into in vivo behavior of TAT agents. Ultimately, this approach aims to provide a robust analytical framework for advancing TAT research and improving clinical translation of alpha-emitting therapeutics.

\section*{Acknowledgements}
TRIUMF receives federal funding via a contribution agreement through the National Research Council of Canada. We also acknowledge added support through the New Frontiers in Research Fund - Exploration NFRFE-2019-00128 and Discovery Grants from the Natural Sciences and Engineering Research Council of Canada, (NSERC): SAPIN-2021-00030 (P. Kunz) and RGPIN-2022-03887 (H. Yang). Many thanks to the Life Science Laboratory team.

\section*{References}
\bibliography{biblio} 
\end{document}